\documentclass[twocolumn,letterpaper,amsmath,amssymb,floatfix,aps,superscriptaddress]{revtex4}

\usepackage{graphicx}
\usepackage{dcolumn}
\usepackage{bm}
\usepackage[usenames]{color}
\usepackage{hyperref}
\usepackage{ulem}

\begin{document}

\title{Iso-Flux Tension Propagation Theory of Driven Polymer Translocation: 
The Role of Initial Configurations}

\author{Jalal Sarabadani}
\email{jalal.sarabadani@aalto.fi}
\affiliation{Department of Applied Physics and COMP Center of Excellence, Aalto University School of Science, 
P.O. Box 11000, FI-00076 Aalto, Espoo, Finland}

\author{Timo Ikonen}
\affiliation{Department of Applied Physics and COMP Center of Excellence, Aalto University School of Science, 
P.O. Box 11000, FI-00076 Aalto, Espoo, Finland}
\affiliation{VTT Technical Research Centre of Finland, P.O. Box 1000, FI-02044 VTT, Finland}

\author{Tapio Ala-Nissila}
\affiliation{Department of Applied Physics and COMP Center of Excellence, Aalto University School of Science, 
P.O. Box 11000, FI-00076 Aalto, Espoo, Finland}
\affiliation{Department of Physics,
Brown University, Providence, Rhode Island 02912-1843}

\begin{abstract}
We investigate the dynamics of pore-driven polymer translocation by 
theoretical analysis and molecular dynamics (MD) simulations. 
Using the tension propagation theory within the constant 
flux approximation we derive an explicit equation of motion for the tension front.
From this we derive a scaling relation for the average translocation time $\tau$, which 
captures the asymptotic result $\tau \propto N_0^{1+\nu}$, where $N_0$ is the chain length and $\nu$ is
the Flory exponent. In addition, we derive the leading correction-to-scaling term to $\tau$ and show that all
terms of order $N_0^{2\nu}$ exactly cancel out, leaving only a finite-chain length correction term due to the
effective pore friction, which is linearly proportional to $N_0$.
We use the model to numerically include fluctuations in the initial 
configuration of the polymer chain in addition to thermal noise. We show that when the {\it cis} side
fluctuations are properly accounted for,
the model not only reproduces previously known results but also considerably improves the estimates of 
the monomer waiting time distribution and the time evolution of the translocation coordinate $s(t)$, showing excellent 
agreement with MD simulations. 
\end{abstract}

\maketitle

\section{Introduction} \label{introduction}

Polymer translocation has in less that 20 years become one of the most active research areas in soft matter biological physics.
Since the initial experimental work of Kasianowicz {\it et al.} \cite{kasi1996} on RNA translocation 
through $\alpha$-hemolysin channels, the interest in the potential technological applications such as
gene therapy, drug delivery and rapid DNA sequencing has motivated a steady flow of experimental and 
theoretical research~\cite{meller2003,Muthukumar_book,Milchev_JPCM,Tapio_review,schadt2010, branton2008,%
storm2005, chuang2001,kantor2004, sung1996,%
metzler2003, muthu1999, sakaue2007,sakaue2008,sakaue2010,saito2011, rowghanian2011,luo2008, saito2012a,%
saito2012b, grosberg2006, luo2009, dubbeldam2011, dubbeldam2007,bhatta2009,bhatta2010,lehtola2009, lehtola2010,%
ikonen2012a,ikonen2012b,ikonen2013}. Of particular interest is the case of the pore-driven polymer translocation, where 
the segment of the polymer inside the pore is driven by an electric field. Unlike the case of unbiased translocation, 
where the polymer supposedly has enough time to equilibrate in some limits 
\cite{unbiased_Slater_1,unbiased_Slater_2, unbiased_Slater_3, unbiased_Polson}
the driven translocation problem is inherently a far-from-equilibrium 
process~\cite{bhatta2009,bhatta2010,lehtola2009,lehtola2010}.

In the recent years, significant advance has been made in the theoretical basis of driven polymer translocation. 
It is now understood that the dynamics of driven translocation is dominated by the drag of the {\it cis} side chain, 
with leading order corrections stemming from the friction of the pore~\cite{ikonen2012a,ikonen2012b,ikonen2013}, with 
the {\it trans} side suspected to have only a minor effect on the whole process~\cite{ikonen2012b,dubbeldam2014,suhonen2014}.
To evaluate the contribution from the {\it cis} side drag, one must study the non-equilibrium time evolution of the chain 
configurations. The basic picture is that of two domains, with the chain divided into mobile and immobile parts, 
where only the segments belonging into the mobile part contribute to the drag. In the simplest description, the process 
can be viewed as a sequential straightening of loops, where the loops between a given segment and the pore need to be 
pulled straight before the segment can experience the force and become mobile~\cite{grosberg2006, dimarzio1979, ikonen2013}. 
Based on this picture, the scaling form $\tau \propto N_0^\alpha$ for the average translocation time $\tau$ as a function 
of chain length $N_0$ can be derived with scaling arguments~\cite{ikonen2013}, giving $\tau(N_0) = c_1 N_0^{1+\nu} +c_2 \tilde{\eta}_p N_0$,
where $c_1$ and $c_2$ are constants. 
Here the first term is due to the {\it cis} side drag and contains the Flory exponent $\nu$ that characterizes the initial 
shape of the chain, given by the end-to-end distance $R\propto N_0^\nu$. The latter term is due to the interaction of the pore 
and the polymer, the strength of which is given by the effective pore friction $\tilde{\eta}_p$.

Thermal fluctuations from the solvent introduce both undulations to the shape of the chain and randomness into the effective 
driving force. Using blob theory, it is possible to describe the shape of the mobile part and the propagation of the 
boundary between the mobile and immobile parts self-consistently~\cite{sakaue2007,sakaue2008,sakaue2010,saito2011,%
rowghanian2011, dubbeldam2011,saito2012a,saito2012b,ikonen2012a,ikonen2012b}. Asymptotic analysis of this 
{\it tension propagation} theory also gives the long chain limit of the translocation time as $\tau= c_1 N_0^{1+\nu}$, similar 
to the simple scaling arguments~\cite{ikonen2013}. Numerical analysis has shown that the finite chain length effects due 
to the pore friction persist for extremely long chains, and that they are responsible for the scatter in the reported values 
of the scaling exponent $\alpha$~\cite{ikonen2012a,ikonen2012b,ikonen2013}. 

With numerical methods, one may also consider the effect of thermal fluctuations to the driving force. Previous results 
indicate that the randomness of the effective force alone is insufficient to explain the fluctuations observed in 
molecular dynamics simulations~\cite{ikonen2012b}. Saito and Sakaue have proposed that for large driving forces the 
uncertainty in the initial configurations would determine the distribution of the translocation time~\cite{saito2012a}.

In this paper, our main aim is to investigate the influence of the uncertainty in the initial chain configuration to translocation 
dynamics by introducing stochasticity to the initial chain configuration at the {\it cis} side. This is based on using the 
Brownian dynamics - tension propagation (BDTP) framework introduced in
Refs.~\cite{ikonen2012a,ikonen2012b}. We modify this approach by deriving the tension propagation 
(TP) equations by assuming a constant monomer flux on the mobile part of the chain in the {\it cis} side. This formalism
leads to an explicit equation of motion for the velocity of the tension front and eliminates 
the need of the original BDTP model to include an approximate initial velocity profile to ensure the conservation mass. 
In addition, the model allows us to derive a finite-size scaling form for the average translocation time, which
is in agreement with ansatz of Ref.~\cite{ikonen2013}.

This paper is organized as follows: In Sec.~\!\ref{model} we demonstrate how to model driven translocation based on the 
iso-flux Brownian dynamics tension propagation (IFTP) formalism. Section~\!\ref{scaling} is devoted to deriving the finite-size scaling form
for the translocation time. In Sec.~\!\ref{initialconfigurations} it is shown how the initial configurations can be incorporated
into the theory. Secs.~\!\ref{average_translocation_time}, \ref{waiting_time}, \ref{distribution_translocation_time} and 
\ref{translocation_coordinate} present the results on the average of the translocation time, waiting time distribution, 
distribution of the translocation time and time evolution of the translocation coordinate, respectively.
Finally, the conclusions and discussion are in Sec.~\!\ref{conclusions}.

\section{Model} \label{model}

For brevity, we use dimensionless units denoted by tilde 
as $\tilde{Y} \equiv Y / Y_u$, with the units of time $t_u \equiv \eta a^2 / (k_B T)$, 
length $s_u \equiv a$, velocity $v_u \equiv a/t_u = k_B T/(\eta a)$, force $f_u \equiv k_B T/a$, 
friction $\Gamma_u \equiv \eta$, and monomer flux $\phi_u \equiv k_B T/(\eta a^2)$,
where $k_B$ is the Boltzmann constant, $T$ is the temperature of the system, $a$ is the segment 
length, and $\eta$ is the solvent friction per monomer.

\begin{figure*}[t]\begin{center}
    \begin{minipage}[b]{0.32\textwidth}\begin{center}
        \includegraphics[width=0.95\textwidth]{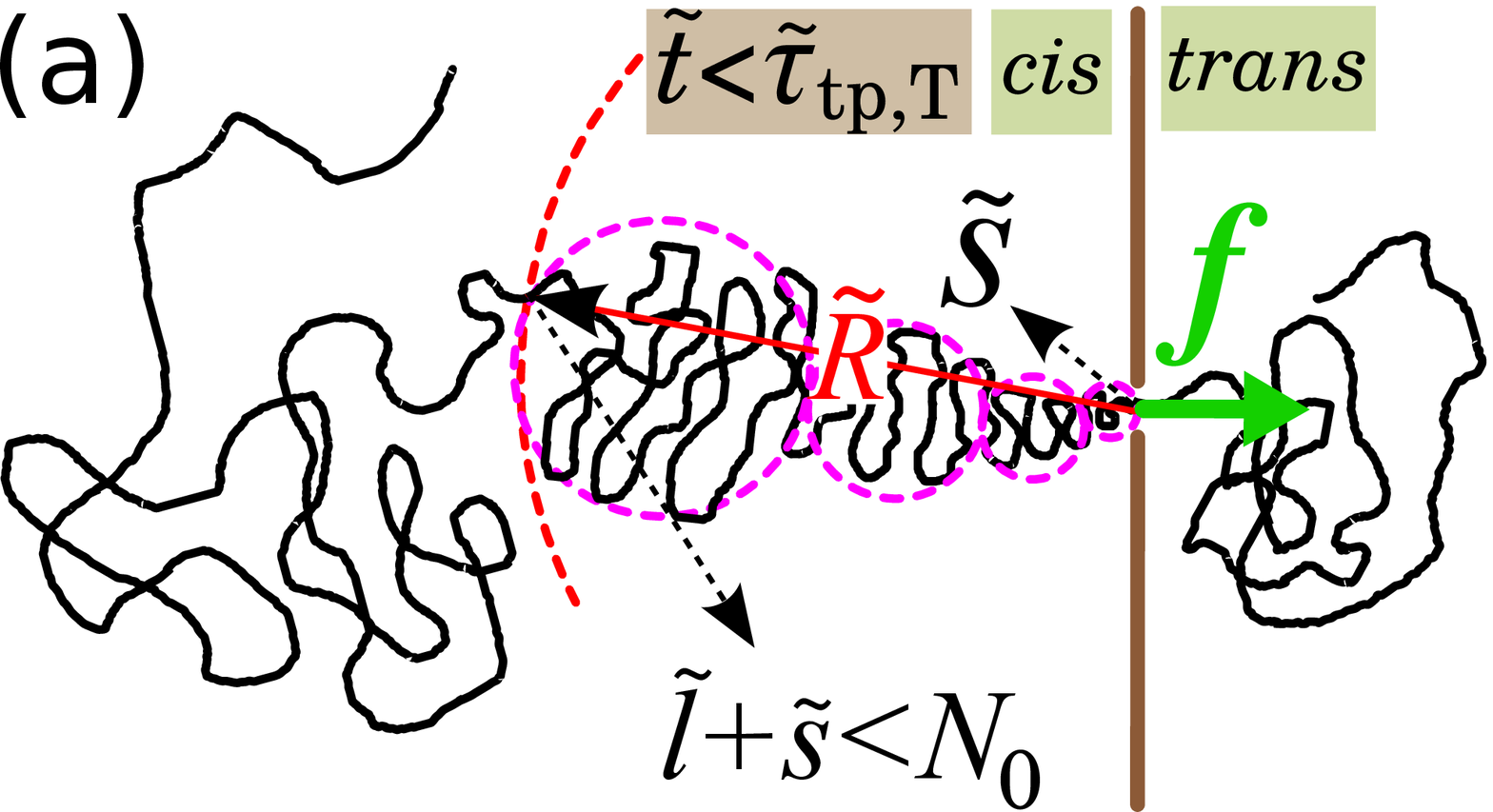}
    \end{center}\end{minipage} \hskip+0cm
    \begin{minipage}[b]{0.32\textwidth}\begin{center}
        \includegraphics[width=0.9\textwidth]{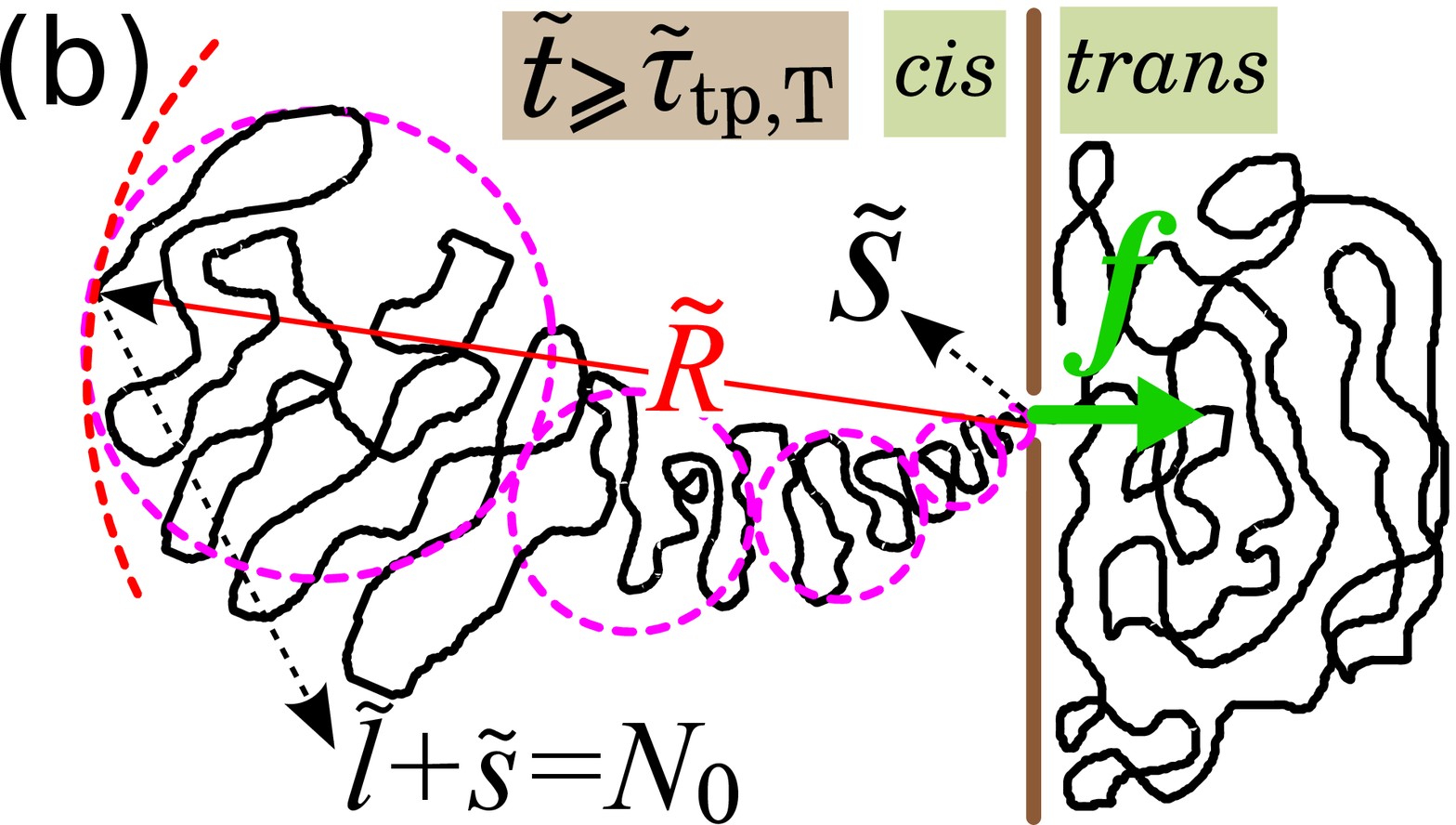}
    \end{center}\end{minipage} \hskip0cm
    \begin{minipage}[b]{0.32\textwidth}\begin{center}
        \includegraphics[width=0.95\textwidth]{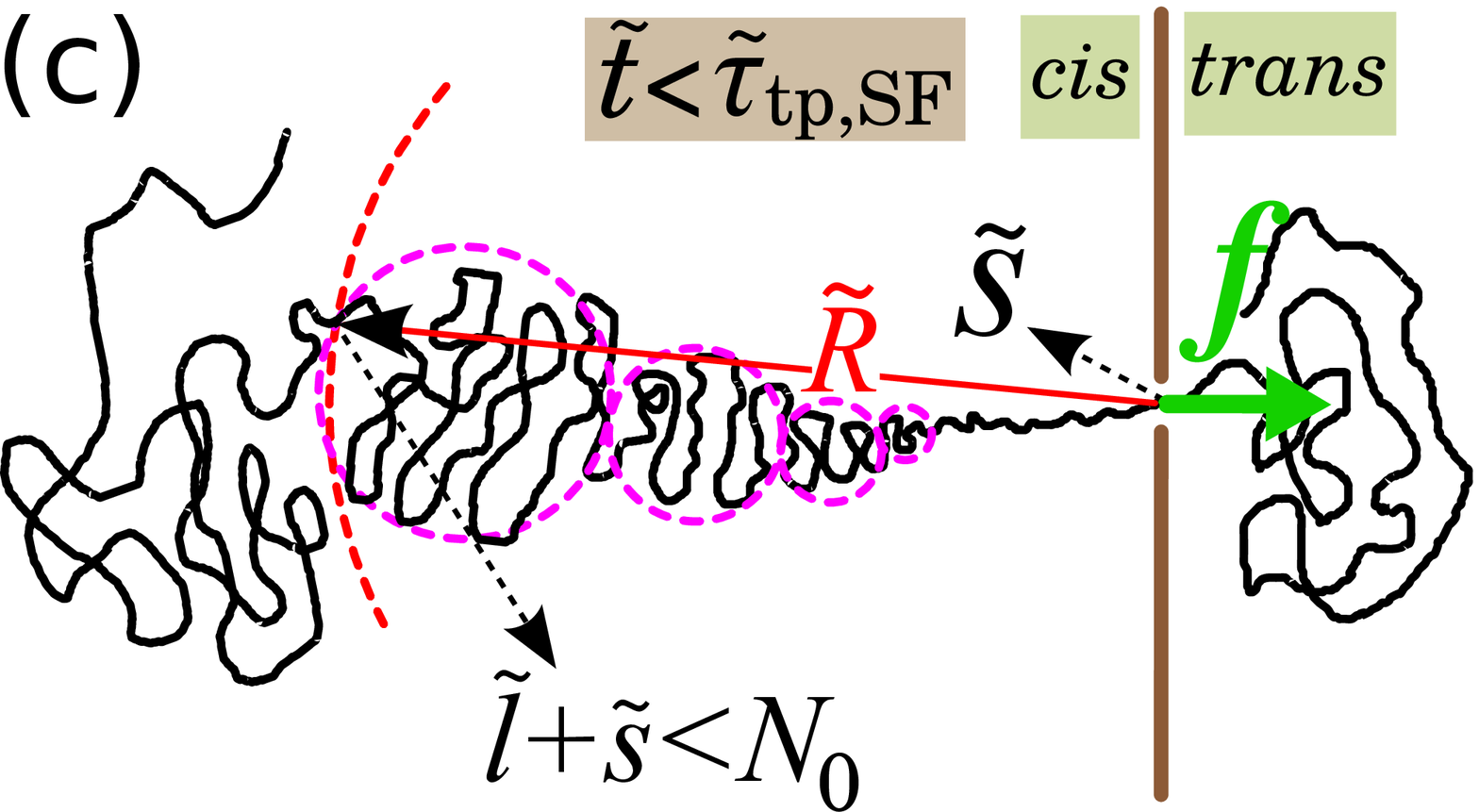}
    \end{center}\end{minipage} \hskip0cm
\caption{(a) A schematic picture of the translocation process during propagation stage for the trumpet regime. 
The driving force $f$ acts on polymer at the pore towards the {\it trans} side. The length of polymer is 
$N_0$ and the number of beads that have already been translocated into the {\it trans} side is denoted by 
$\tilde{s}$. The number of beads influenced by the tension in the {\it cis} side is $\tilde{l} + \tilde{s}$ 
which is less than the number of total beads in the polymer $N_0$ during propagation stage. The location 
of the last blob is determined by $\tilde{R}$. (b) The translocation process when the tension front reaches 
the chain end and after it for the trumpet regime (post propagation stage). (c) The same as (a) but for the
stem-flower regime. $\tilde{\tau}_{\textrm{tp,T}}$ and $\tilde{\tau}_{\textrm{tp,SF}}$ define the propagation 
times in the trumpet and stem-flower regimes, respectively, as in Eq.~\!(\ref{propagation_time_1}).} 
\label{fig:schimatic}
\end{center}
\end{figure*}

As a basic framework we use Brownian dynamics (BD) in the overdamped limit, similar to Refs.~\cite{ikonen2012a,ikonen2012b}. 
The BD equation is written for the translocation coordinate $\tilde{s}$ that gives the length of the chain on the {\it trans} 
side. The equation reads as 
\begin{equation}
\tilde{\Gamma} (\tilde{t}) \frac{d \tilde{s}}{ d \tilde{t}} =
(1- \gamma ') \bigg[ \frac{1}{N_0 - \tilde{s}} - \frac{1}{\tilde{s}} \bigg] 
+ \tilde{f} + \tilde{\zeta} (\tilde{t}) \equiv  \tilde{f}_{\textrm{tot}},
\label{BD_equation}
\end{equation}
where $\tilde{\Gamma}$ is the effective friction, and $\tilde{\xi} (\tilde{t})$ is Gaussian white noise
which satisfies $\langle \xi (t) \rangle = 0$ and $\langle \xi (t) \xi (t) \rangle = 2 \Gamma 
k_B T \delta (t - t ')$, $\gamma '$ is the surface exponent ($\gamma '= 0.5$ for ideal chains, and $\approx ~ 0.95, 
\approx~ 0.69$ for self-avoiding chains in two and three 
dimensions, respectively), $N_0$ is the total number of beads in the chain (the contour length of 
the chain is $L= a N_0$), $\tilde{f}$ is the external driving force and $\tilde{f}_{\textrm{tot}}$ 
is the total force. The effective friction $\tilde{\Gamma}$ depends on the pore friction $\tilde{\eta}_p$ and the 
drag force on the {\it cis} side. As the dynamical {\it trans} side contribution to the dynamics has been shown to be
insignificant~\cite{ikonen2012a,ikonen2012b,ikonen2013,dubbeldam2014,suhonen2014}, we absorb it into the constant 
pore friction $\tilde{\eta}_p$. The dynamics of the {\it cis} side is solved with the TP equations.

To derive the TP equations, we use arguments similar to Rowghanian {\it et al.}~\cite{rowghanian2011}. We assume 
that the flux $\tilde{\phi}\equiv d\tilde{s}/d\tilde{t}$ of monomers on the mobile domain of the {\it cis} side 
and through the pore is constant in space, but evolves in time. The boundary between the mobile and immobile 
domains, the tension front, is located at distance $\tilde{x}=-\tilde{R}(\tilde{t})$ from the pore. Inside the 
mobile domain, the external driving force is mediated by the chain backbone from the pore at $\tilde{x}=0$ all 
the way to the last mobile monomer $N$ located at the tension front. The magnitude of the tension force at 
distance $\tilde{x}$ can be calculated by considering the force-balance relation for the differential element 
$d \tilde{x}$ that is located between $\tilde{x}$ and $\tilde{x} + d\tilde{x}$. By integrating the force-balance 
relation over the distance
from the pore entrance to $\tilde{x}$, the tension force can be obtained as $\tilde{f}(\tilde{x},\tilde{t}) = 
\tilde{f}_0 - \tilde{\phi} (\tilde{t}) \tilde{x}$ (see Appendix \ref{f_of_x} for details). Here $ \tilde{f}_0  
\equiv \tilde{f}_{\textrm{tot}} - \tilde{\eta}_p\tilde{\phi}(\tilde{t})$ is the force at the pore entrance.

Closer to the tension front the mediated force is therefore smaller, as it is diminished by the drag of all 
the preceding monomers. According to blob theory, the chain then assumes a trumpet-like shape with the narrow 
end closer to the pore, such as shown in Figs.~\ref{fig:schimatic}~\!(a) and (b). For a moderate external driving 
force, i.e. $N_0^{-\nu} \ll \tilde{f}_0 \ll 1$, the monomer density at the pore is greater than unity, and the shape 
of the chain resembles a trumpet. This is classified as a {\it trumpet} (TR) regime.  For a stronger external 
driving force, $1 \ll \tilde{f}_0 \ll N_0^{\nu}$, the force is large enough to completely straighten a small 
part of the chain. This part is called the stem, while the part following it is called the flower, corresponding 
to the {\it stem-flower} (SF) regime (see Fig.~\ref{fig:schimatic}~\!(c)). In both regimes the tension front is
located at the farthest blob from the pore as depicted in Fig.~\ref{fig:schimatic}. 

Integration of the force balance equation over the mobile domain gives an expression for the monomer flux as a 
function of the force and the linear size of the mobile domain as
\begin{equation}
\tilde{\phi} (\tilde{t}) = \frac{\tilde{f}_{\textrm{tot}} (\tilde{t})}
{\tilde{\eta}_{\textrm{p}} + \tilde{R} (\tilde{t}) }.
\label{phi_equation}
\end{equation}
Equation \eqref{BD_equation} and the definition of the flux, $\tilde{\phi}\equiv d\tilde{s}/d\tilde{t}$, can be 
then used to find the expression for the effective friction as
\begin{equation}
\tilde{\Gamma} (\tilde{t}) = \tilde{R}(\tilde{t}) + \tilde{\eta}_p. 
\label{Gamma_equation}
\end{equation}

Equations \eqref{BD_equation}, \eqref{phi_equation} and \eqref{Gamma_equation} determine the time evolution of $\tilde{s}$, 
but the full solution still requires the knowledge of $\tilde{R}(\tilde{t})$. The derivation of the equation of motion 
of $\tilde{R}(\tilde{t})$ is done separately for the {\it propagation} and {\it post propagation} stages. In the propagation stage, 
the tension has not reached the final monomer in Fig.~\ref{fig:schimatic}~\!(a). Here the propagation of the tension 
front into the immobile domain is determined by the geometric shape of the immobile domain. In practice, one uses the 
scaling relation of the end-to-end distance of the self-avoiding chain to arrive at the closure relation 
$\tilde{R} = A_{\nu} N^{\nu}$, where $A_\nu$ is a constant prefactor and $N$ is the last monomer inside the tension front. 
As shown in Appendix \ref{app_equation_for_R}, one can then derive an equation of motion for the tension front as
\begin{equation}
\dot{\tilde{R}} (\tilde{t}) =
\frac{\nu A_{\nu}^{ \frac{1}{\nu} }  \tilde{R} (\tilde{t})^{ \frac{\nu -1}{\nu} }
\big[  ({\cal{L}}_{\textrm{a}} + {\cal{G}}_{\textrm{a}}) \times
\dot{\tilde{f} }_{\textrm{tot}}  (\tilde{t}) + \tilde{\phi} (\tilde{t}) \big] }
{ 1 + \nu A_{\nu}^{ \frac{1}{\nu} }  \tilde{R} (\tilde{t})^{ \frac{\nu -1}{\nu} }  
{\cal{L}}_{\textrm{a}} \times \tilde{\phi} (\tilde{t}) },
\label{evolution_of_R_propagation}
\end{equation}
where ${\cal{L}}_{\textrm{a}}$ and ${\cal{G}}_{\textrm{a}}$ are functions of $\tilde{\phi}$, $\tilde{\eta}_p$ and $\nu$,
$\dot{\tilde{f} }_{\textrm{tot}}$ is the time derivative of $\tilde{f}_{\textrm{tot}}$, and the subscript "$\textrm{a}$" in 
${\cal{L}}_{\textrm{a}}$ and ${\cal{G}}_{\textrm{a}}$ stands for the trumpet regime as $\textrm{T}_+$ and $\textrm{T}_-$ 
correspond to positive and negative values of $\tilde{\phi}$ respectively, and for the stem-flower regime as $\textrm{SF}$. 

In the post propagation stage in Fig.~\ref{fig:schimatic}~\!(b), every monomer on the {\it cis} is affected by the tension. 
Therefore, we have the condition $N=N_0$. Since $N$ is also equal to the number of monomers already translocated, $\tilde{s}$, 
plus the number of monomers currently mobile on the {\it cis} side, $\tilde{l}$, the correct closure relation for the post 
propagation stage is $\tilde{l}+\tilde{s}=N_0$. The equation of motion for the tension front is then derived as
\begin{equation}
\dot{\tilde{R}} (\tilde{t}) = \frac{ ({\cal{L}}_{\textrm{a}} + {\cal{G}}_{\textrm{a}}) 
\dot{\tilde{f} }_{\textrm{tot}}  (\tilde{t}) + \tilde{\phi} (\tilde{t})}
{ \tilde{\phi} (\tilde{t}) \times {\cal{L}}_{\textrm{a}} },
\label{evolution_of_R_post_propagation_2}
\end{equation}
which is demonstrated in Appendix \ref{app_equation_for_R}.

The self-consistent solution for the model in the propagation stage can be obtained from Eqs.~\!(\ref{BD_equation}), 
(\ref{phi_equation}), (\ref{Gamma_equation}) and (\ref{evolution_of_R_propagation}). Correspondingly, in the post 
propagation one uses the set of Eqs.~\!(\ref{BD_equation}), (\ref{phi_equation}), 
(\ref{Gamma_equation}) and (\ref{evolution_of_R_post_propagation_2}).

\section{Scaling of translocation time} \label{scaling}

To obtain some analytical results, it is useful to consider the approximation of constant force, $\tilde{f}_{\textrm{tot}} = 
\tilde{f}$. Then Eq.~\!\eqref{phi_equation} reduces to $\tilde{\phi}(\tilde{t})= \tilde{f}/\left( \tilde{R}(\tilde{t}) + \tilde{\eta}_p \right)$.
In the stem-flower and trumpet regimes, the number of mobile monomers on the {\it cis} side is given by $\tilde{l}_{\textrm{SF}} = \tilde{R} + 
C_\nu \tilde{\phi}^{-1}$, and $\tilde{l}_{\textrm{T}} = \frac{\nu}{2 \nu -1} \tilde{\phi}^{(\nu -1)/\nu} \tilde{R}^{(2 \nu -1)/\nu}$, 
respectively, where $C_\nu = (1-\nu)/(2\nu -1)$. This together with the conservation of mass, $N= \tilde{s}+ \tilde{l}$, 
allows one to solve the propagation time $\tilde{\tau}_\mathrm{tp}$ by integration of $N$ from $0$ to $N_0$. The result is
\begin{equation}
\tilde{\tau}_\mathrm{tp,a} = \frac{1}{\tilde{f}} \bigg[ \int_0^{N_0} \tilde{R}(N)dN + \tilde{\eta}_p N_0 \bigg]
- \Delta \tilde{\tau}_\mathrm{a},
\label{propagation_time_1}
\end{equation}
where the subscript "$\textrm{a}$" denotes SF and T, and
\begin{align}
\Delta \tilde{\tau}_\mathrm{SF}= &
\left( \frac{1}{\tilde{f}} +\frac{C_\nu}{\tilde{f}^2} \right) \left[ \frac{1}{2} \tilde{R}^2(N_0) 
+ \tilde{\eta}_p \tilde{R}(N_0) \right], \notag\\
\Delta \tilde{\tau}_\mathrm{T}= 
& \tilde{f}^{-\frac{1}{\nu}}\bigg[ \int_{0}^{\tilde{R}(N_0)} d \tilde{R} ~ \tilde{R}^{1-\frac{1}{\nu}} 
\big( \tilde{R} + \tilde{\eta}_{\textrm{p}} \big)^{\frac{1}{\nu}} \notag\\
& + C_\nu \int_{0}^{\tilde{R}(N_0)} d \tilde{R} ~ \tilde{R}^{2-\frac{1}{\nu}} 
\big( \tilde{R} + \tilde{\eta}_{\textrm{p}} \big)^{\frac{1}{\nu} -1} \bigg].
\label{propagation_time_2}
\end{align}

In the post propagation stage, one sets the condition $dN/d\tilde{t} = 0$ and integrates $\tilde{R}$ from $\tilde{R}(N_0)$ to $0$. 
The result for the post-propagation time $\tilde{\tau}_\mathrm{pp}$ is
\begin{equation}
\tilde{\tau}_\mathrm{pp,a} = \Delta \tilde{\tau}_\mathrm{a}.
\end{equation}

The time over the whole translocation process is then given by
\begin{align}
\tilde{\tau}_{\textrm{a}} = \tilde{\tau}_\mathrm{tp,a} + \tilde{\tau}_\mathrm{pp,a} & = 
\frac{1}{\tilde{f}}\left[ \int_0^{N_0} \tilde{R}(N)dN + \tilde{\eta}_p N_0 \right] \notag \\
& = \frac{A_\nu}{(1+\nu)\tilde{f}} N_0^{1+\nu}+ \frac{\tilde{\eta}_p}{\tilde{f}} N_0 .
\end{align}
This is a remarkable result in the sense that although terms proportional to $N_0^{2\nu}$ appear in the intermediate steps, 
as predicted for instance in Refs.~\!\cite{saito2012a,dubbeldam2011}, the terms are canceled out in the expression for 
the total translocation time. This result is in agreement with the previously proposed scaling analysis and MD simulations 
in Ref.~\cite{ikonen2013}.

\section{Distribution of Initial Configurations} \label{initialconfigurations}

In previous works with the BDTP model, an average end-to-end distance $\tilde{R} = A_{\nu} N^{\nu}$ was used with a constant 
coefficient $A_{\nu}= 1.15$ in 3D~\cite{ikonen2012a,ikonen2012b}. To study the influence of initial configurations 
on the translocation process we employ a new probability distribution function to sample the 
end-to-end distance of the chain. To obtain the distribution, we have done Langevin-thermostatted molecular dynamics 
simulations of self-avoiding chains tethered onto an impenetrable wall and calculated the end-to-end distance of 
chain. We simulated several chain lengths up to $N_0=321$, with standard Kremer-Grest bead-spring model of the 
chain and other parameters typically used in the MD simulations. For detailed account of the simulation method and 
the parameters, see e.g. Refs.~\cite{ikonen2012a, ikonen2012b}.

\begin{figure}[t]\begin{center}
\includegraphics[scale=0.32]{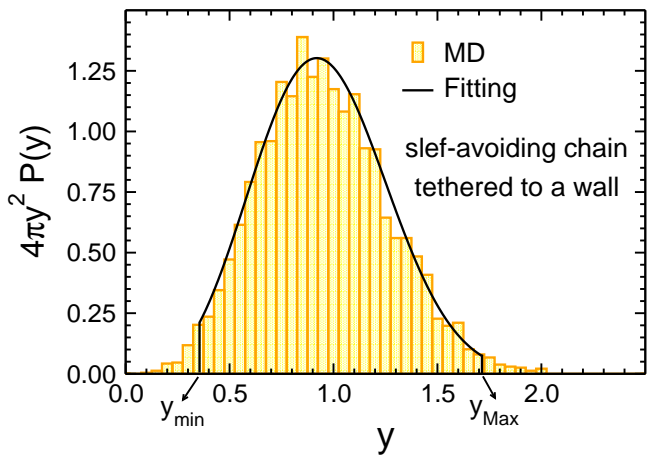}
\caption{MD data for the probability distribution function multiplied by $4 \pi y^2$ 
(yellow bars), with fitting to the MD data shown as a black line. The fitting curve is 
$4 \pi y^2 P(y)$ where $ P(y)  = 4 \pi y^2 A ~\! y^B {\textrm{exp}} \big[ Cy^D ]$ and
$A=0.4252$, $B=1.0310$, $C=-1.4417$, $D=2.6203$. 
}
\label{probability_end_to_end}
\end{center}
\end{figure}

The distribution of the end-to-end distances for $N_0=321$ is shown in Fig.~\!\ref{probability_end_to_end}. 
An analytical function was fitted to the cumulative distribution function constructed from the data by minimizing 
the squared error with the conditions that the total probability and the second moment are equal to unity. 
The obtained analytical probability distribution function can be written as
\begin{equation}
P(y)  = A ~\! y^B {\textrm{exp}} \big[ Cy^D ],
\label{distribution_initial_configuration}
\end{equation}
where $A=0.4252$, $B=1.0310$, $C=-1.4417$, $D=2.6203$, and $y$ is the normalized end-to-end distance 
$y= \tilde{R} / \langle \tilde{R} \rangle $. The fitted function was also compared to MD data of shorter chains 
($N_0=81$ and $N_0=161$) with a Kolmogorov-Smirnov test, showing that within 99~\% statistical confidence the 
shorter chains follow the same distribution for the normalized end-to-end distance.

Using Eq.~\!(\ref{distribution_initial_configuration}) one can sample over many different initial configurations. 
By choosing $y$ from the probability distribution function in Eq.~\!(\ref{distribution_initial_configuration}) and
redefining $\tilde{R}$ as $\tilde{R} = A_{\nu} (y) ~ N_0^{\nu} $, one can incorporate an approximate distribution of 
$\tilde{R}$ into the TP theory through $A_{\nu} (y) = y A_{\nu}$. For numerical reasons we have covered the range
$y_{\textrm{min}} < y < y_{\textrm{max}}$, where $y_{\textrm{min}}= 0.356$ and $y_{\textrm{max}}= 1.718$. This is 
justified  because 97 $\%$ of the area below the curve in Fig.~\!\ref{probability_end_to_end} is still covered 
by choosing these cutoffs.

\section{Results} \label{results}

\subsection{Average translocation time} \label{average_translocation_time}

The most fundamental property related to the translocation process is the average translocation time $\tilde{\tau}$. 
According to the analysis of 
Section~\ref{scaling}, the translocation time $\tilde{\tau}$ depends on the chain length $N_0$ as
\begin{equation}
\tilde{\tau} \equiv c_1 (\tilde{f}, \nu, A_{\nu}) ~ N_0^{1+\nu} + c_2 (\tilde{f}) ~ \tilde{\eta}_{\textrm{p}} N_0.
\label{trans_time_two_terms}
\end{equation}
Written in the conventional scaling form, $\tilde{\tau}\propto N_0^\alpha$, it is evident that the effective exponent $\alpha$ 
is a function of chain length due to the correction-to-asymptotic-scaling term in Eq.~\eqref{trans_time_two_terms}.

\begin{figure*}[t]
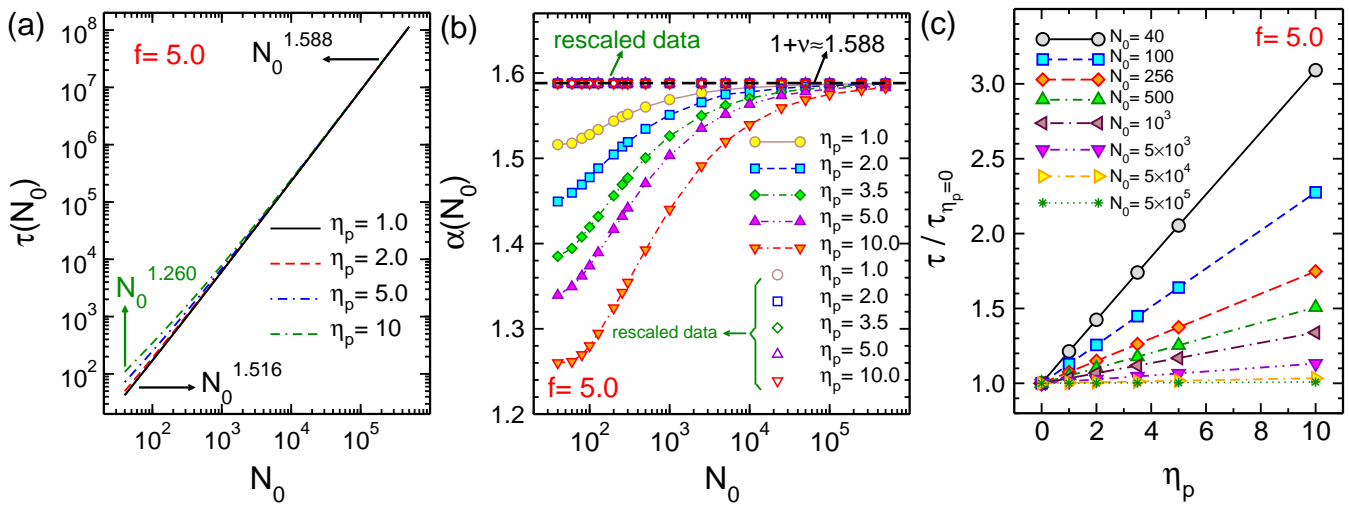
\begin{center}
    \begin{minipage}[b]{0.317\textwidth}\begin{center}
        \includegraphics[width=0.9999\textwidth]{figure3a.eps}
    \end{center}\end{minipage} \hskip-0.15cm
    \begin{minipage}[b]{0.365\textwidth}\begin{center}
        \includegraphics[width=0.962\textwidth]{figure3b.eps}
    \end{center}\end{minipage} \hskip-0.15cm
    \begin{minipage}[b]{0.317\textwidth}\begin{center}
        \includegraphics[width=0.9999\textwidth]{figure3c.eps}
    \end{center}\end{minipage} \hskip-0.1cm
\caption{(a) The translocation time as a function of the chain length, $N_0$, for fixed values of 
the force, $f= 5.0$, and $A_{\nu}= 1.15$ for various values of $\eta_{\textrm{p}}$. 
The effective exponent for the shortest chain, $N_0 = 40$, and pore friction 
$\eta_{\textrm{p}}= 1.0$ is $1.516$, while for highest value of pore friction $\eta_{\textrm{p}}= 10.0$ it is 
$1.260$. The effective exponent for the longest chain, $N_0 = 5 \times 10^{5}$ is
$1.588$. (b) The effective exponent $\alpha (N_0)$ as a function of the chain length for various values of pore 
friction $\eta_{\textrm{p}}$, and the rescaled exponent that is also plotted as a function of chain 
length for various $\eta_{\textrm{p}}$. As can be seen, the rescaled exponent curves for different 
values of $\eta_{\textrm{p}}$ collapse on a single master curve, i.e. $\alpha^{\dag} (N_0) = 1 + \nu$, as denoted by 
rescaled data in the figure. (c) The normalized translocation time, $\tau / \tau_{\eta_{\textrm{p}=0}}$, 
plotted as a function of pore friction, $\eta_{\textrm{p}}$, for various chain lengths.} 
\label{trans_time_figure}
\end{center}
\end{figure*}

To illustrate this behavior, we have solved the model numerically. For parameter values $f=5.0$, $k_BT=1.2$, 
$\eta=0.7$, and pore frictions $\eta_{\textrm{p}}= 1.0, 2.0, 5.0$ and 10.0, the translocation time as a function 
of chain length is shown in Fig.~\!\ref{trans_time_figure}~\!(a). Here we have used a fixed value $A_\nu=1.15$ 
and we have set the stochastic term $\zeta$ in the force to zero in order to be able to simulate chain lengths up to 
$N_0\approx 10^6$. For short chains, there is  a clear dependence in the slope on the pore friction. For the long 
chains, this dependence dies off as the asymptotic limit of $\alpha = 1 + \nu$ is reached.

The dependence on the pore friction is even more clear in Fig.~\!\ref{trans_time_figure}~\!(b), where we have plotted 
the effective translocation exponent defined as $\alpha (N_0) = \textrm{d} \ln \tau/( \textrm{d} \ln N_0 )$ \cite{ikonen2013} 
for different values of pore friction. 
We have checked the chain length dependence of the translocation exponent for the specific case of fixed pore friction 
$\eta_{\textrm{p}}= 3.5$, when the thermal fluctuations as well
as distribution of the initial configurations of the chain are taken into account. For both cases the translocation 
exponents are the same as of the deterministic case within a quite good accuracy.
As mentioned, the dependence of the translocation exponents on the pore friction is more 
pronounced for short chain lengths. To show that the difference from the asymptotic value is indeed caused 
by the pore friction term, and not some other finite size effects, we define a rescaled translocation time as
\begin{equation}
\tilde{\tau}^{\dag} = \tilde{\tau} - c_2 \tilde{\eta}_{\textrm{p}} N_0 = c_1 N_0^{1+\nu} \sim N_0^{\alpha^{\dag}} ,
\label{trans_time_rescaled_as_chain_length_eq}
\end{equation}
where $\alpha^{\dag} \equiv 1 + \nu$ is the rescaled translocation exponent which does not depend on the
chain length. As explained in Ref.~\cite{ikonen2013}, $c_1$ and $c_2$ can be obtained by calculating 
the intercept and slope
of the curve $\tau / N_0^{1+\nu}$ as a function of $\tilde{\eta}_{\textrm{p}} N_0^{-\nu}$, respectively.
Calculating the rescaled exponent as $\alpha^{\dag} (N_0) = \textrm{d} \ln \tau^{\dag}/( \textrm{d} \ln N_0 )$ 
it is found that it is indeed equal to $1 + \nu$ for all chain lengths, 
independent of pore friction which is demonstrated in Fig.~\!\ref{trans_time_figure}~\!(b). 
This result 
is in excellent agreement with the molecular dynamics simulation results discussed in Ref.~\cite{ikonen2013}.

To further illustrate the influence of the pore friction on the translocation time, in Fig.~\!\ref{trans_time_figure}~\!(c) 
the normalized translocation time, $\tau / \tau_{\eta_{\textrm{p}} = 0}$, has been plotted as a function of the 
pore friction, $\eta_{\textrm{p}}$, for various values of chain length, $N_0 = 40 - 5 \times 10^5$. As it 
can be seen the normalized translocation time is influenced strongly by the pore friction for shorter chains
while for longest chain the translocation time is constant for different values of the pore friction.

\subsection{Waiting time distribution} \label{waiting_time}

\begin{figure}[t]\begin{center}
\includegraphics[scale=0.33]{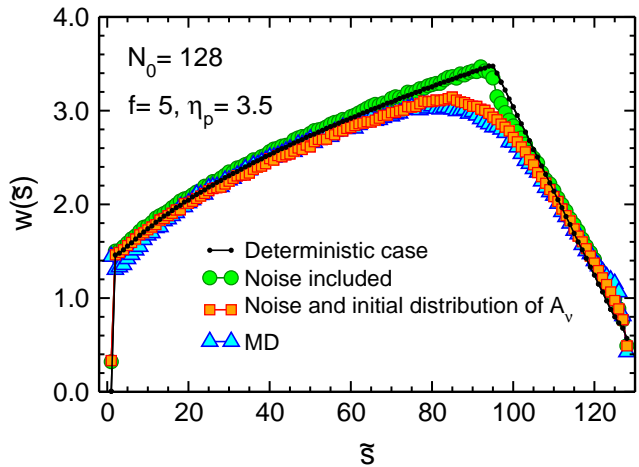}
\caption{Waiting time, $w (\tilde{s})$, as a function of the translocation coordinate, $\tilde{s}$. Here, we 
present waiting time for different cases when both of the force and $A_{\nu}= 1.15$ are deterministic 
(black curve), force is chosen randomly but $A_{\nu}= 1.15$ is deterministic (green circles), both the 
force and $A_{\nu}$ are stochastic (red squares), and finally MD simulation data 
(blue triangles).}
\label{waiting_time_distribution_figure}
\end{center}
\end{figure}

An important quantity in examining the dynamics of the translocation process is the monomer 
waiting time, which is defined as the time that
each monomer or segment spends at the pore during the translocation process. The waiting time is calculated for 
each monomer, and averaged over the different simulation trajectories. Here we have calculated the waiting time as 
a function of the translocation coordinate $\tilde{s}$ and present it in Fig.~\!\ref{waiting_time_distribution_figure} 
for a fixed chain length $N_0 = 128$, external driving force $f = 5.0$ and $\eta_{\textrm{p}} = 3.5$. 
It can be seen that the translocation process is a far-from-equilibrium process and has two different stages. First 
one is the propagation stage where as the time passes more monomers are moved and involved in the drag friction 
force. Therefore the friction increases monotonically until it gets its maximum value which happens when the 
tension reaches the chain end. The second stage of the translocation process is called the post propagation stage 
that starts when the tension reaches the chain end. During this stage the remaining part of the chain in the 
{\it cis} side is sucked through the pore and at the end the translocation process ends when the whole chain
passes through the pore to the {\it trans} side. 

We can now use the IFTP model to separately examine the influence of thermal fluctuations in the noise
and the distribution of the initial configuration of the chain.  The results 
are shown in Fig.~\!\ref{waiting_time_distribution_figure}. The black curve corresponds to the deterministic case, where both 
the force $f = 5.0$ and the amplitude $A_{\nu}= 1.15$ are fixed. The green circles
show the waiting time when the force includes the stochastic component (noise) $\zeta$ and $A_{\nu}= 1.15$ is fixed. 
As can be seen, the mean values are almost identical to the first deterministic case. The 
red squares exhibit the waiting time when both the force and $A_{\nu}$ are stochastic, i.e. the force includes noise 
and the initial distribution of $A_\nu$ is sampled from Eq.~\!(\ref{distribution_initial_configuration}). 
The main effect of the stochastic sampling of the initial configurations 
is to smoothen the transition from the propagation to the post-propagation stage. This is a feature that is also seen 
in molecular dynamics simulations (blue triangles), where the initial configuration is sampled by thermalizing the 
polymer before each simulation trajectory. All in all, there is now a very good agreement between the theory and the MD 
simulations.

\subsection{Distribution of translocation time} \label{distribution_translocation_time}

\begin{figure}[b]\begin{center}
\includegraphics[scale=0.325]{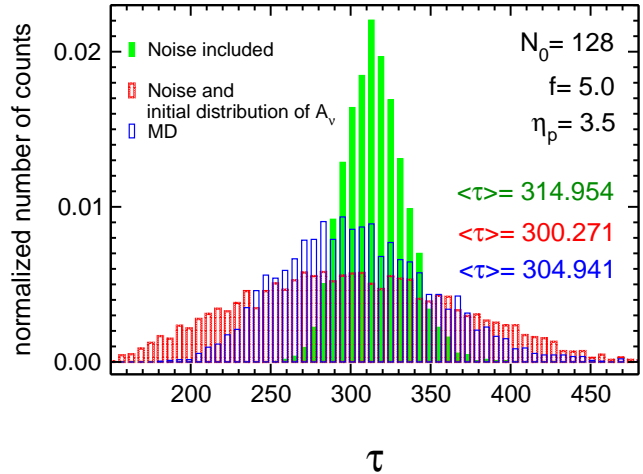}
\caption{ The translocation time histogram as a function of translocation time $\tau$. 
The green bars present the normalized histogram when $A_{\nu}= 1.15$ is deterministic 
while the external driving force is $f=5.0$ and the total force includes the stochastic contribution. The red bars 
correspond to solutions where $A_{\nu} (y)$ is also chosen from 
Eq.~\!(\ref{distribution_initial_configuration}). The histogram of the translocation time based on 
MD simulation is illustrated by blue bars.}
\label{trasn_time_histo}
\end{center}
\end{figure}

Another quantity which is of fundamental interest is the translocation time distribution which is depicted in 
Fig.~\!\ref{trasn_time_histo}. The green bars present the histogram of the translocation time for fixed 
$A_{\nu}= 1.15$ (noise included). Here the distribution is solely due to the randomness of the driving force. 
The red bars show the histogram where $A_{\nu} (y)$ has been sampled using 
Eq.~\!(\ref{distribution_initial_configuration}) as $A_{\nu} (y) = y A_{\nu}$ where $A_{\nu} = 1.15$ (noise and 
initial distribution of $A_\nu$). To compare the results of the theoretical model with MD data, the histogram 
of the translocation times based on MD simulations is also shown as blue bars. As it can be seen, the distribution 
with fixed $A_\nu$ is much narrower than the MD result. This is in agreement with the observations of 
Ref.~\cite{ikonen2012a}. However, there is a much better agreement with MD when the initial configurations are randomly 
sampled. Here the distribution gets wider and agrees quite well with the MD data, in particular for long
translocation times. However, the model predicts slightly faster translocation events than the MD. The 
reason for this is easy to understand. In choosing the prefactor $A_\nu$ as the parameter describing the variance 
in the initial configurations, we ensure that the end-to-end distance distribution is well reproduced. However, 
the shape of the chain remains unchanged. Specifically, the form $\tilde{R} \propto N_0^\nu$ excludes 
configurations where the chain extends far away from the pore but loops back so that the end-to-end distance 
becomes small. Thus the drag due to the long loops is not entirely accounted for, and the effective friction 
and consequently the translocation time are underestimated. This result also indicates that it may be necessary 
to express the equilibrium shape of the chain with more than just one parameter to capture the variation in the 
translocation time in detail.

\subsection{Evolution of the translocation coordinate $\tilde{s}$ as a function of time} 
\label{translocation_coordinate}

\begin{figure*}[t]
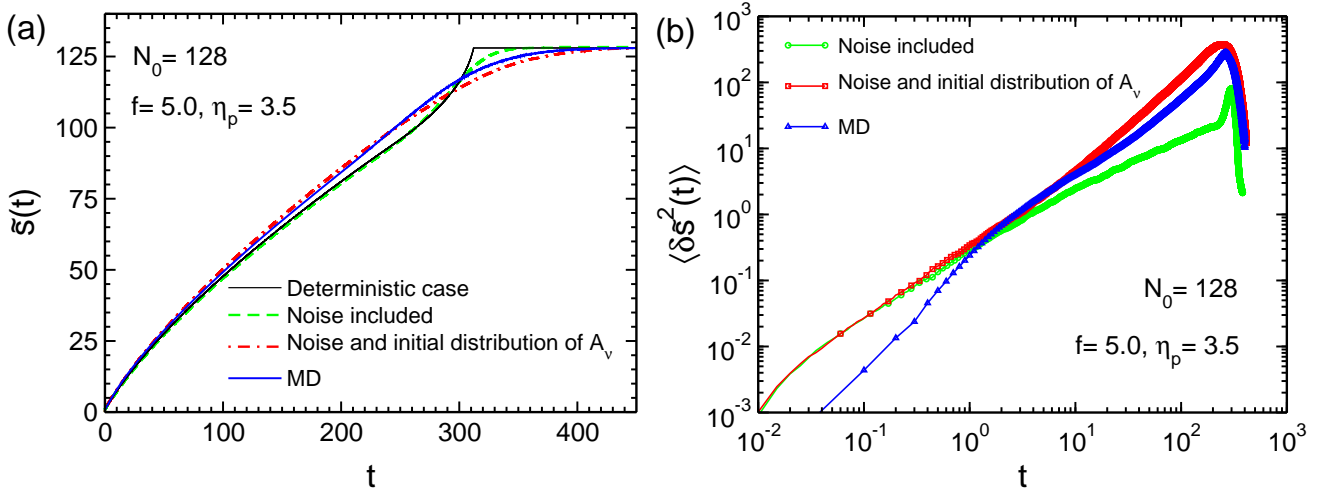
\begin{center}
    \begin{minipage}[b]{0.475\textwidth}\begin{center}
        \includegraphics[width=0.99\textwidth]{figure6a.eps}
    \end{center}\end{minipage} \hskip+0.0cm
    \begin{minipage}[b]{0.49\textwidth}\begin{center}
        \includegraphics[width=0.99\textwidth]{figure6b.eps}
    \end{center}\end{minipage} \hskip0cm
\caption{(a) The translocation coordinate, $\tilde{s} (t)$, as a function of time, $t$, when both the force 
and $A_{\nu}$ are deterministic (black solid line), force includes noise but $A_{\nu}$ is deterministic 
(green dashed line), both force and $A_{\nu}$ are stochastic (red dashed-dotted line), and the MD data (blue line). 
(b) The fluctuations of the translocation coordinate, $\langle \delta \tilde{s}^2 (t) \rangle \equiv \langle \tilde{s}^2 
(t) \rangle - \langle \tilde{s} (t) \rangle^2$, as a function of time for the cases when the force includes noise 
while $A_{\nu}$ is deterministic (green), both force and $A_{\nu}$ are stochastic
(red), and for MD simulations (blue). Here, we have chosen fixed chain length $N_0=128$, external 
driving force $f~\!=~\!5.0$ and the pore friction as $\eta_{\textrm{p}}=3.5$.} 
\label{s_as_func_of_time}
\end{center}
\end{figure*}

Finally, we examine how the translocation coordinate and its fluctuations evolve in time. These quantities
could not be explained with the previous BDTP theory of Refs. \cite{ikonen2012a,ikonen2012b}. Here we have again 
chosen the chain length $N_0=128$,  driving force $f=5.0$ 
and the pore friction as $\eta_{\textrm{p}}=3.5$. The results for $\tilde{s}(t)$ can be seen in 
Fig.~\!\ref{s_as_func_of_time}~\!(a), and for the variance $\langle \delta \tilde{s}^2 (t) \rangle \equiv 
\langle \tilde{s}^2 (t) \rangle - \langle \tilde{s} (t) \rangle^2$ in Fig.~\!\ref{s_as_func_of_time}~\!(b). 
We have again solved the model with the stochastic 
force term first off and with fixed initial configuration (black curve), then with thermal noise included in the 
force (green curves), and both thermal noise and randomly sampled initial configurations (red curves). 
We also compare the results with MD, shown with blue curves.

The fully deterministic solution (the black $\tilde{s}(t)$ curve) is quite different from the MD solution 
towards the end, and approaches the final value of $\tilde{s}=128$ much more sharply. The shape is very 
similar to that shown in, e.g., Refs.~\cite{sakaue2010, dubbeldam2014}. Adding the fluctuations to the 
driving force makes the approach to the terminal value a bit smoother. However, the larger difference 
comes again from the initial configurations. With the random selection of the end-to-end distance, the 
results match very well with MD data.

For the fluctuations of $\tilde{s}$, the results are similar. With just the thermal fluctuations in the 
driving force, the variance increases much slower than the MD results. This is consistent with the earlier 
study of Ref.~\cite{ikonen2012b}. When the initial configuration is randomized, the results are much 
improved and are again in good agreement with MD. However, similar to the distribution of the translocation 
time discussed above, the magnitude of the fluctuations is slightly overestimated.

\section{Conclusions} \label{conclusions}

In this paper we have derived a model of driven polymer translocation based on combined Brownian dynamics-tension 
propagation theory in the constant flux approximation. The model gives an explicit equation of motion for the position 
of the tension front and allows a full characterization of the translocation process. In particular, it can be used to 
derive a finite-size formula for the scaling of the translocation time as a function of the chain length, revealing that 
the main correction-to-scaling term comes from the pore friction and is linearly proportional to $N_0$. The model
reproduces the chain length dependence of the effective scaling exponents from the previous BDTP theory 
\cite{ikonen2012a,ikonen2012b,ikonen2013}. Moreover, it allows a detailed study of the interplay between thermal
noise in the force and initial distribution of the chain configurations. The analysis presented here shows that by including the
latter effect, quantities such as the waiting time, the distribution of translocation time and the dynamics and fluctuations in
the translocation coordinate are in good agreement with the MD data. This reveals the important role of the {\it cis} side
of the chain to driven translocation and justifies the approximation to neglect the {\it trans} side degrees of freedom
from the model.

\begin{acknowledgments}
This work was supported by the Academy of Finland through its Centres
of Excellence Program (2012-2017) under Project No.~\!915804.
\end{acknowledgments}

\appendix 

\section{Force at distance $\tilde{x}$ to the pore}\label{f_of_x}

The value of the force as a function of the distance to the pore, $\tilde{x}$, on the {\it cis}
side can be obtained for the trumpet regime by integrating the force balance relation,
$d \tilde{f} (\tilde{x}') = - \tilde{\phi} (\tilde{t}) ~ d~\!\!\tilde{x}'$,
for a differential element $d \tilde{x}'$ over the distance between 0 and $\tilde{x}$ as
\begin{equation}
\tilde{f} (\tilde{x}) = \tilde{f}_0 - \tilde{\phi} (\tilde{t}) \tilde{x},
\label{force_as_func_x}
\end{equation}
where $\tilde{f}_0$ is the force at the entrance of the pore. Note that here we have used the 
iso-flux assumption which means that the value of the monomer flux, $\tilde{\phi}$, is constant 
over the integration range $[0, \tilde{x}]$.

In the stem-flower regime, the region of mobile beads is separated 
into two sub-regions. In the stem region the chain is straightened because 
the tension force is stronger and in the flower region as the tension force is weaker, blobs are formed. 
The border between the stem and the flower regions is at $\tilde{x}=\tilde{r} (\tilde{t})$ where the tension 
force has the value of unity. 
Writing the force balance equation for a differential element and integrating over the stem region, 
$\tilde{r} (\tilde{t})$ can be found as 
\begin{equation}
\tilde{r} (\tilde{t}) = \frac{\tilde{f}_0 - 1}{\tilde{\phi} (\tilde{t})}.
\label{force_stem}
\end{equation}
Then by integrating the force balance equation over the distance between $\tilde{r}$ and $\tilde{x}$, 
that $\tilde{f} (\tilde{r}) = 1$, in the flower regime one can write the following relation 
\begin{equation}
\tilde{f} (\tilde{x}) = 1 - \tilde{\phi} (\tilde{t}) ~\! (\tilde{x} - \tilde{r}).
\label{force_flower}
\end{equation}
Combining Eqs.~\!(\ref{force_stem}) and (\ref{force_flower}) the same relation similar to the trumpet 
regime can be obtained for the stem-flower regime as 
$\tilde{f}(\tilde{x}) = \tilde{f}_0 - \tilde{\phi} (\tilde{t}) \tilde{x}$.

\section{Equation of motion for the tension front}\label{app_equation_for_R}

To find an equation for the time evolution of the tension front location, $\tilde{R}$, for the propagation stage
one must calculate and then substitute $\tilde{l}$, the number of mobile beads on the {\it cis} side, and 
$\tilde{s}$ into the closure relation 
\begin{equation}
\tilde{R} = A_{\nu} [\tilde{l}+\tilde{s}]^{\nu},
\label{closure_relation_el_s}
\end{equation}
and then perform the time derivative of $\tilde{R}$ that can be read as a function of 
$\dot{\tilde{l}} (\tilde{t})$ and $\dot{\tilde{s}} (\tilde{t})$ as 
\begin{equation} %2
\dot{\tilde{R}} (\tilde{t}) = 
\nu A_{\nu}^{1/\nu} \tilde{R} (\tilde{t}) ^{\frac{\nu -1}{\nu}}
[\dot{\tilde{l}} (\tilde{t}) + \dot{\tilde{s}} (\tilde{t})],
\label{R_dot}
\end{equation}
where by definition 
\begin{equation} %2
\frac{d \tilde{s} (\tilde{t})}{d \tilde{t}} = \dot{\tilde{s}} (\tilde{t}) = \tilde{\phi} 
(\tilde{t}).
\label{s_dot}
\end{equation}

The number of mobile monomers in the {\it cis} side, i.e. $\tilde{l}(\tilde{t})$, 
is obtained by integrating the linear monomer number density, $\tilde{\sigma} (\tilde{t})$, over the 
distance between $0$ and $\tilde{R}$. Therefore, first the monomer number density must be obtained.
To this end the blob theory can be used. When a blob is constructed by applying the tension force 
on the backbone of the chain, the blob size, $\tilde{\xi} (\tilde{x})$, can be obtained as 
$\tilde{\xi} (\tilde{x}) = 1/|\tilde{f} (\tilde{x})|$ where $\tilde{f} (\tilde{x}) = \tilde{f}_0 - 
\tilde{\phi} (\tilde{t}) \tilde{x}$ is the force at the distance $\tilde{x}$ to the pore in the 
{\it cis} side which has been obtained in Appendix \ref{f_of_x}.
On length scales shorter than the Pincus blob size, $\tilde{\xi} (\tilde{x})$, the chain behaves as if undisturbed 
by the external driving force and the blob size scales as $\tilde{\xi} = g^{\nu}$, where $g$ is the 
number of monomers inside the blob. Finally the monomer number density is given by 
$\tilde{\sigma} (\tilde{x}, \tilde{t}) = \frac{g}{\tilde{\xi}} = \tilde{\xi}^{\frac{1}{\nu} -1} = 
\big|\tilde{f}(\tilde{x}) \big|^{1-\frac{1}{\nu}} $. 
Using the above monomer number density the number of mobile monomers 
in the {\it cis} side can be derived as
\begin{equation}
\tilde{l}(\tilde{t}) = \int_{0}^{\tilde{R} (\tilde{T}) } \tilde{\sigma} (\tilde{x}, \tilde{t}) d \tilde{x}.
\label{def_el}
\end{equation}
Therefore, for the trumpet regime
\begin{eqnarray}
\tilde{l}_{\textrm{T}} (\tilde{t})  &=& \int_{0}^{\tilde{R}(\tilde{t})}
\tilde{\sigma} (\tilde{x}, \tilde{t}) d\tilde{x} =
\int_{0}^{\tilde{R}(\tilde{t})}
|\tilde{f} (\tilde{x})|^{(\nu-1)/\nu} d\tilde{x} = \nonumber\\
&=& \int_{0}^{\tilde{R}(\tilde{t})}
\big|\tilde{f}_0 - \tilde{\phi} (\tilde{t}) \tilde{x}\big|^{(\nu-1)/\nu} d\tilde{x} \nonumber\\
&=&
\int_{0}^{\tilde{R}(\tilde{t})}
\bigg| \tilde{\phi} (\tilde{t}) \tilde{R} (\tilde{t}) 
- \tilde{\phi} (\tilde{t}) \tilde{x}\bigg|^{(\nu-1)/\nu} d\tilde{x}.
\label{el_trumpet_1}
\end{eqnarray}
Consequently
\begin{subequations}
\begin{align}
\tilde{l}_{\textrm{T}_{+}}(\tilde{t}) &= \frac{\nu}{(2\nu-1)}
\tilde{\phi} (\tilde{t})^{\frac{\nu-1}{\nu}} \tilde{R} (\tilde{t})^{\frac{2\nu-1}{\nu}}
~~~:\tilde{\phi} (\tilde{t})>0 ,
\\
\tilde{l}_{\textrm{T}_{-}}(\tilde{t}) &= \frac{\nu}{(2\nu-1)}  
[-\tilde{\phi} (\tilde{t})]^{\frac{\nu-1}{\nu}} \tilde{R} (\tilde{t})^{\frac{2\nu-1}{\nu}}
\!:\tilde{\phi} (\tilde{t})<0,
\end{align} 
\label{el_trumpet_2}
\end{subequations}
where the subscript $\textrm{T}$ denotes the trumpet regime, and $+$ and $-$ stand for the positive and negative 
values of $\tilde{\phi} (\tilde{t})$, respectively.

To obtain $\tilde{l}_{\textrm{SF}}(\tilde{t})$, which is the number of mobile monomers in the {\it cis} side 
in the stem-flower regime, similar to the procedure for the trumpet regime, one has to integrate the linear monomer 
number density, $\tilde{\sigma} (\tilde{t})$, over the distance from 0 to $\tilde{R}$, i.e.
\begin{eqnarray}
\tilde{l}_{\textrm{SF}}  (\tilde{t})  &=& \int_{0}^{\tilde{R}(\tilde{t})}
\tilde{\sigma} (\tilde{x}, \tilde{t}) d\tilde{x}  \nonumber\\
&=& \int_{0}^{\tilde{r}(\tilde{t})}
\tilde{\sigma} (\tilde{x}, \tilde{t}) d\tilde{x} +
\int_{\tilde{r}(\tilde{t})}^{\tilde{R}(\tilde{t})}
\tilde{\sigma} (\tilde{x}, \tilde{t}) d\tilde{x}  \nonumber\\
&=& \tilde{r}(\tilde{t})
+ \int_{\tilde{r}(\tilde{t})}^{\tilde{R}(\tilde{t})}
|\tilde{f} (\tilde{x})|^{(\nu-1)/\nu} d\tilde{x}  \nonumber\\
&=&
\frac{\tilde{\phi} (\tilde{t}) \tilde{R} (\tilde{t}) - 1}
{\tilde{\phi} (\tilde{t})} \nonumber\\
&& + 
\int_{\tilde{r}(\tilde{t})}^{\tilde{R}(\tilde{t})}
\bigg| \tilde{\phi} (\tilde{t}) \tilde{R} (\tilde{t}) 
- \tilde{\phi} (\tilde{t}) \tilde{x}\bigg|^{(\nu-1)/\nu} d\tilde{x}.
\label{stem_flower_el}
\end{eqnarray}
Performing the integral yields $\tilde{l}_{\textrm{SF}}(\tilde{t})$ as
\begin{equation}
\tilde{l}_{\textrm{SF}}(\tilde{t}) = \tilde{R} (\tilde{t}) + \frac{1-\nu}{(2\nu-1)} 
\frac{1}{\tilde{\phi} (\tilde{t})},
\label{el_stem_flower_2}
\end{equation}
where the index $\textrm{SF}$ denotes the stem-flower regime. 
Then the time derivative of the number of mobile beads, $\dot{\tilde{l}} (\tilde{t})$ can be cast into 
\begin{equation}
\dot{\tilde{l}}_{\textrm{a}} (\tilde{t}) = {\cal{L}}_{\textrm{a}} \times
[\dot{\tilde{f} }_{\textrm{tot}}  (\tilde{t})  - \tilde{\phi} (\tilde{t}) \dot{\tilde{R} } (\tilde{t}) ]
+ \tilde{{\cal{G}}}_{\textrm{a}} \times \dot{\tilde{f} }_{\textrm{tot}}  (\tilde{t}),
\label{el_dot}
\end{equation}
where ${\textrm{a}}={\textrm{T}_{+}},~{\textrm{T}_{-}}~{\textrm{and}}~{\textrm{SF}}$, and 
\begin{subequations}
\begin{align}
{\cal{L}}_{\textrm{T}_{+}} &= \frac{ 1 }{ \tilde{\eta}_{\textrm{p}} + \tilde{R} (\tilde{t}) }
\bigg\{ -\frac{\nu}{(2\nu-1) \tilde{\phi} (\tilde{t})^2 } 
\big[ \tilde{\phi} (\tilde{t}) \tilde{R} (\tilde{t}) \big]^{\frac{2\nu-1}{\nu}} \nonumber\\
&- \frac{  \tilde{\eta}_{\textrm{p}} }{  \tilde{\phi} (\tilde{t})  } 
\big[ \tilde{\phi} (\tilde{t}) \tilde{R} (\tilde{t})  \big]^{\frac{\nu-1}{\nu}}
\bigg\}, 
\hspace{+1.5cm}:~\tilde{\phi} (\tilde{t}) >0
\\
{\cal{L}}_{\textrm{T}_{-}} &= \frac{1}{\tilde{\eta}_{\textrm{p}} + \tilde{R} (\tilde{t}) } 
\bigg\{   \frac{\nu}{(2\nu-1) \tilde{\phi} (\tilde{t})^2 } 
\big[ -\tilde{\phi} (\tilde{t}) \tilde{R} (\tilde{t}) \big]^{\frac{2\nu-1}{\nu}} \nonumber\\
&- \frac{  \tilde{\eta}_{\textrm{p}} }{  \tilde{\phi} (\tilde{t})  } 
\big[ - \tilde{\phi} (\tilde{t}) \tilde{R} (\tilde{t}) \big]^{\frac{\nu-1}{\nu}} \bigg\},
\hspace{+1.0cm}:~\tilde{\phi} (\tilde{t}) <0
\\
\tilde{{\cal{G}}}_{\textrm{T}_{+}} &= \frac{ 1 }{ \tilde{\phi} (\tilde{t}) } 
\big[ \tilde{\phi} (\tilde{t}) \tilde{R} (\tilde{t}) \big]^{\frac{\nu-1}{\nu}}
\hspace{+1.8cm}:~\tilde{\phi} (\tilde{t}) >0,
\\
\tilde{{\cal{G}}}_{\textrm{T}_{-}} &= \frac{ 1 }{ \tilde{\phi} (\tilde{t}) } 
\big[ - \tilde{\phi} (\tilde{t}) \tilde{R} (\tilde{t}) \big]^{\frac{\nu-1}{\nu}}
\hspace{+1.4cm}:~\tilde{\phi} (\tilde{t}) <0,
\\
{\cal{L}}_{\textrm{SF}} &=
\frac{1}{ \tilde{\eta}_{\textrm{p}} + \tilde{R} (\tilde{t}) } 
\bigg[ -\frac{\tilde{\eta}_{\textrm{p}} \tilde{\phi} (\tilde{t}) 
+ \tilde{\phi} (\tilde{t}) \tilde{R} (\tilde{t}) }
{ \tilde{\phi} (\tilde{t})^2 }  \nonumber\\
&+ \frac{\nu-1}{(2\nu-1) \tilde{\phi} (\tilde{t})^2 } 
\bigg] , \\
{\cal{G}}_{\textrm{SF}} &= \frac{1}{\tilde{\phi} (\tilde{t}) }.
\end{align} 
\label{L_G}
\end{subequations}

Combining Eqs.~(\ref{R_dot}), (\ref{s_dot}) and (\ref{el_dot}) the equation for 
the time evolution of the tension front can be written as
\begin{equation}
\dot{\tilde{R}} (\tilde{t}) =
\frac{\nu A_{\nu}^{ \frac{1}{\nu} }  \tilde{R} (\tilde{t})^{ \frac{\nu -1}{\nu} }
\big[  ({\cal{L}}_{\textrm{a}} + {\cal{G}}_{\textrm{a}}) \times
\dot{\tilde{f} }_{\textrm{tot}}  (\tilde{t}) + \tilde{\phi} (\tilde{t}) \big] }
{ 1 + \nu A_{\nu}^{ \frac{1}{\nu} }  \tilde{R} (\tilde{t})^{ \frac{\nu -1}{\nu} }  
{\cal{L}}_{\textrm{a}} \times \tilde{\phi} (\tilde{t}) }.
\label{evolution_of_R_propagation_appendix}
\end{equation}

In the post propagation stage, the closure relation is given by the sum over the number 
of mobile beads in the {\it cis} side, $\tilde{l}$, and the number of translocated beads, 
$\tilde{s}$, as $\tilde{l} + \tilde{s} = N_0$. The time derivative of this closure relation 
is 
\begin{equation}
\dot{\tilde{l}} + \dot{\tilde{s}} = 0.
\label{time_derivative_closure_post_propag}
\end{equation}

Combining Eqs.~\!(\ref{s_dot}), (\ref{el_dot}) and (\ref{time_derivative_closure_post_propag})
the equation of motion for the tension front in the post propagation stage can be cast into:
\begin{equation} %2
\dot{\tilde{R}} (\tilde{t}) = \frac{ ({\cal{L}}_{\textrm{a}} + {\cal{G}}_{\textrm{a}}) 
\dot{\tilde{f} }_{\textrm{tot}}  (\tilde{t}) + \tilde{\phi} (\tilde{t})}
{ \tilde{\phi} (\tilde{t}) \times {\cal{L}}_{\textrm{a}} }.
\label{evolution_of_R_post_propagation_2_Appendix}
\end{equation}

\end{document}